\documentclass{emulateapj}

\newcommand{\kms}{\hbox{km s$^{-1}$}}
\newcommand{\msun}{\hbox{$M_\odot$}}
\newcommand{\lsun}{\hbox{$L_\odot$}}

\newcommand{\re}{\hbox{$R_{\rm e}$}}

\newcommand{\se}{\hbox{$\sigma_{\rm e}$}}

\newcommand{\refsec}[1]{Section~\ref{#1}}
\newcommand{\reffig}[1]{Fig.~\ref{#1}}
\newcommand{\refeq}[1]{equation~(\ref{#1})}

\slugcomment{Received 2009 June 2; accepted 2009 September 4}
\shorttitle{Dynamical masses of early-type galaxies at \lowercase{$z\sim2$}}
\shortauthors{Cappellari et al.}
\begin{document}

\title{Dynamical masses of early-type galaxies at \lowercase{$z\sim2$}:\\ Are they truly superdense?\altaffilmark{1}}

\author{Michele Cappellari,\altaffilmark{2}
    S.~di Serego Alighieri,\altaffilmark{3}
    A.~Cimatti,\altaffilmark{4}
    E.~Daddi,\altaffilmark{5}
    A.~Renzini,\altaffilmark{6}
    J.~D.~Kurk,\altaffilmark{7}
    P.~Cassata,\altaffilmark{8}
    M.~Dickinson,\altaffilmark{9}
    A.~Franceschini,\altaffilmark{10}
    M.~Mignoli,\altaffilmark{11}
    L.~Pozzetti,\altaffilmark{11}
    G.~Rodighiero,\altaffilmark{10}
    P.~Rosati,\altaffilmark{12}
    G.~Zamorani\altaffilmark{11}
}

\altaffiltext{1}{Based on observations collected at the European Southern Observatory, Paranal, Chile, ESO Large Programs 173.A–0687}
\altaffiltext{2}{Sub-Department of Astrophysics, University of Oxford, Denys Wilkinson Building, Keble Road, OXFORD, OX1 3RH, ENGLAND}
\altaffiltext{3}{INAF - Osservatorio Astrofisico di Arcetri, Largo E. Fermi 5, 50125 Firenze, Italy}
\altaffiltext{4}{Dipartimento di Astronomia, Universit\`a di Bologna, Via Ranzani 1, 40127 Bologna, Italy}
\altaffiltext{5}{CEA – Saclay, DSM/DAPNIA/Service d'Astrophysique, 91191 Gif-sur-Yvette Cedex, France}
\altaffiltext{6}{INAF - Osservatorio Astronomico di Padova, Vicolo dell'Osservatorio 5, 35122 Padova, Italy}
\altaffiltext{7}{Max-Planck-Institut f\"ur Astronomie, K\"onigstuhl 17, 69117 Heidelberg, Germany}
\altaffiltext{8}{Department of Astronomy, University of Massachusetts, LGRT-B 619E, 710 North Pleasant Street, Amherst, MA 01003-9305, USA}
\altaffiltext{9}{NOAO – Tucson, 950 North Cherry Avenue, Tucson, AZ 85719, USA}
\altaffiltext{10}{Universit\`a di Padova, Dipartimento di Astronomia, Vicolo dell'Osservatorio 2, 35122 Padova, Italy}
\altaffiltext{11}{INAF - Osservatorio Astronomico di Bologna, Via Ranzani 1, 40127 Bologna, Italy}
\altaffiltext{12}{European Southern Observatory, Karl Schwarzschild Str.~2, 85748 Garching bei M\"unchen, Germany}

\begin{abstract}
We measured stellar velocity dispersions $\sigma$ and derived dynamical masses of 9 massive ($M\approx10^{11}$ \msun) early-type galaxies (ETG) from the GMASS sample at redshift $1.4\la z\la2.0$. The $\sigma$ are based on individual spectra for two galaxies at $z\approx1.4$ and on a stacked spectrum for 7 galaxies with $1.6<z<2.0$, with 202-h of exposure at the ESO Very Large Telescope.  We constructed detailed axisymmetric dynamical models for the objects, based on the Jeans equations, taking the observed surface brightness (from deep HST/ACS observations), PSF and slit effects into account. Our dynamical masses $M_{\rm Jeans}$ agree within $\la30$\% with virial estimates $M_{\rm vir}=5\times\re\sigma^2/G$, although the latter tend to be smaller. Our $M_{\rm Jeans}$ also agrees within a factor $\la2$ with the $M_{\rm pop}$ previously derived using stellar population models and 11 bands photometry. This confirms that the galaxies are intrinsically massive. The inferred mass-to-light ratios $(M/L)_U$ in the very age-sensitive rest frame $U$-band are consistent with passive evolution in the past $\sim1$ Gyr (formation redshift $z_f\sim3$). A `bottom-light' stellar Initial Mass Function (IMF) appears to be required to ensure close agreement between $M_{\rm Jeans}$ and $M_{\rm pop}$ at $z\sim 2$, as it does at $z\sim 0$. The GMASS ETGs are on average more dense than their local counterpart. However a few percent of local ETGs of similar dynamical masses also have comparable $\sigma$ and mass surface density $\Sigma_{50}$ inside \re.
\end{abstract}


\section{Introduction}

In the hierarchical galaxy formation paradigm, where galaxies are assembled by the merging of multiple building blocks in a universe dominated by dark matter \citep[e.g.][]{Springel2005}, the most massive early-type galaxies (ETGs) are assembled last. However observations in the local universe and at high redshift seem to converge towards a `downsizing' mechanism for ETGs formation in which the the stars of the most massive systems formed at the highest redshifts ($z\ga3$), while the stars in the smaller ones were produced over more extended periods of time \citep{Cowie1996,Heavens2004,Thomas2005,Treu2005,Renzini2006}. A way to reconcile the apparent contradiction between these two pictures is to assume that the stars in the massive systems formed via efficient star formation processes at high redshift and were later assembled into larger systems via mostly collisionless mergers \citep{DeLucia2006,Khochfar2006,Naab2009}.

An important test for this scenarios is constituted by the mass and size distribution of ETG at $z\ga1$. Contrary to the expectations the most massive ones appear to be already in place \citep{Cimatti2004,Cimatti2006,Glazebrook2004,Scarlata2007} but have much smaller sizes than their local counterparts \citep{Daddi2005,diSeregoAlighieri2005,Trujillo2006}. This suggests that they might not be the direct precursors of present-day ETGs and mergers must play a role in their evolution \citep{Trujillo2007,Trujillo2009,Zirm2007,Longhetti2007,Toft2007,Cimatti2008,vanDokkum2008,vanderWel2008a,Buitrago2008,Bernardi2009}.

There are however concerns affecting the compactness determinations due to possible observational biases affecting either the mass or size estimate (due to the surface brightness dimming, the presence of AGNs or nuclear starbursts) of galaxies at high redshift. Here we try to to address these concerns by measuring the velocity dispersion $\sigma$ of the stars, related to the density, and deriving masses via dynamical models of mass-selected ETGs at $1.4\la z\la 2.0$. We assumed a flat Universe with $H_0=70$ \kms\ Mpc$^{-1}$, $\Omega_m=0.3$, $\Omega_\Lambda=0.7$.

\section{Velocity dispersion determination}
\label{sec:dispersion}

\subsection{Spectroscopic and photometric data}

The sample under exam comes from the \emph{Galaxy Mass Assembly ultra-deep Spectroscopic Survey} (GMASS)\footnote{http://www.arcetri.astro.it/$\sim$cimatti/gmass/gmass.html} within the redshift range $1.4\la z\la 2.0$ \citep[hereafter C08]{Cimatti2008}. It was flux-selected at 4.5 $\micron$ using the Great Observatories Origin Deep Survey GOODS-South public image taken with IRAC on the Spitzer Space Telescope (Dickinson et al., in preparation).

The GMASS optical multi-slit spectroscopy used here was obtained with the ESO VLT $+$ FORS2 (MXU mode) in the wavelength range 600–-1000 nm with the grism 300I, using very long integration times of up to 32 h per spectroscopic mask, and with a slit width of 1 arcsec. We adopted as instrumental resolution the mean $\sigma_{\rm instr}=130\pm21$ \kms\ of the values derived from sky emission lines and a star, where the error is half the difference between the two determinations. This conservative error also accounts for the small dependence of the resolution with wavelength. We also use public HST/ACS/F850LP ($z$-band) photometry from GOODS-South \citep{Giavalisco2004}.

\subsection{Library of stellar templates}

The FORS2 observations span a rest-frame UV wavelength range of 230--385 nm at the mean redshift $z\approx1.6$ of the GMASS sample. To measure stellar kinematics we need stellar templates in the UV and we cannot use the extensive ground-based stellar libraries. Moreover no empirical UV library span the full required spectral range.

For this we use {\em synthetic} libraries, which now can reproduce spectra of real stars remarkably well \citep{Munari2005,Martins2007}. The mismatch in minor spectral features is not critical when working with low-$S/N$ spectra dominated by systematics. Here we selected as templates a subset of 33 models from the high-resolution $R=20,000$ synthetic spectral library\footnote{http://archives.pd.astro.it/2500-10500/} by \citet{Munari2005} spanning a wide range of temperatures $3500\le T\le10,000$ and surface gravities $0\le\log g\le5$, at solar metallicity and abundance.

\subsection{Individual spectra at $z\approx1.4$}
\label{sec:kinematics_individual}

\begin{table*}
\centering
\caption{Sample of GMASS passive early type galaxies and measured parameters}
\tabcolsep=1.7pt
\begin{tabular}{cccccccccccccc}
\tableline
ID &
$z$ &
$\Delta V$ &
$\sigma_{\rm pred}$ &
$\sigma_\star$ &
$\Delta\sigma_\star$ &
$S/N$ &
$R_{\rm e}$ &
$R_{\rm e}$ &
$\log L_U$ &
$(M/L)_{\rm Jeans}$ &
$\log M_{\rm Jeans}$  &
$\log M_{\rm vir}$ &
$\log M_{\rm pop}$ \\
& & (\kms) & (\kms) & (\kms) & (\kms) & & (arcsec) & (kpc) & ($\lsun_U$) & ($\msun/\lsun_U$)&(\msun)&(\msun)&(\msun)\\
(1) & (2) & (3) & (4) & (5) & (6) & (7) & (8) & (9) & (10) & (11) & (12) & (13) & (14) \\
\tableline
 0472 & 1.9077      & 36 & 191 & --- & ---& 2.7 & 0.06 & 0.54 & 11.04 &  0.38$\pm$0.09\tablenotemark{a} & 10.64\tablenotemark{a} & 10.53\tablenotemark{a} & 10.49 \\
 0996 & 1.3844      & 30 &  98 & --- & ---& 2.4 & 0.13 & 1.06 & 10.36 &   ---                           &  ---  &  ---  & 10.16 \\
 1498 & 1.8491      & 16 & 157 & --- & ---& 2.5 & 0.14 & 1.18 & 10.94 &  0.98$\pm$0.22\tablenotemark{a} & 10.93\tablenotemark{a} & 10.83\tablenotemark{a} & 10.61 \\
 2111 & 1.6102      & 19 & 185 & --- & ---& 4.0 & 0.09 & 0.80 & 10.85 &  0.79$\pm$0.18\tablenotemark{a} & 10.75\tablenotemark{a} & 10.68\tablenotemark{a} & 10.61 \\
 2148 & 1.6118      & 23 & 248 & --- & ---& 6.5 & 0.14 & 1.22 & 11.00 &  0.89$\pm$0.20\tablenotemark{a} & 10.95\tablenotemark{a} & 10.84\tablenotemark{a} & 11.02 \\
 2196 & 1.6063      & 28 & 180 & --- & ---& 3.1 & 0.17 & 1.40 & 10.89 &  0.92$\pm$0.21\tablenotemark{a} & 10.85\tablenotemark{a} & 10.89\tablenotemark{a} & 10.79 \\
 2239 & 1.4149      & 15 & 113 & 111 & 35 & 4.5 & 0.25 & 2.09 & 10.60 &  1.23$\pm$0.78                  & 10.69 & 10.52 & 10.54 \\
 2286 & 1.6020      & 29 & 135 & --- & ---& 3.2 & 0.18 & 1.48 & 10.72 &  1.73$\pm$0.39\tablenotemark{a} & 10.95\tablenotemark{a} & 10.91\tablenotemark{a} & 10.56 \\
 2355 & 1.6097      & 32 & 127 & --- & ---& 2.2 & 0.12 & 1.00 & 10.78 &   ---                           &  ---  &  ---  & 10.36 \\
 2361 & 1.6096      & 18 & 197 & --- & ---& 4.1 & 0.15 & 1.26 & 10.86 &  0.93$\pm$0.21\tablenotemark{a} & 10.83\tablenotemark{a} & 10.85\tablenotemark{a} & 10.83 \\
 2470 & 1.4149      & 11 & 157 & 141 & 26 & 7.6 & 0.18 & 1.53 & 10.92 &  0.66$\pm$0.24                  & 10.74 & 10.61 & 10.71 \\
 2543 & 1.6149      & 54 & 141 & --- & ---& 1.8 & 0.22 & 1.88 & 10.60 &   ---                           &  ---  &  ---  & 10.69 \\
 2559 & 1.9816      & 28 & 147 & --- & ---& 2.4 & 0.19 & 1.61 & 10.91 &   ---                           &  ---  &  ---  & 10.67 \\
\tableline
New Stack & $1.6<z<2.0$ & 15 & 205\tablenotemark{b} & 202 & 23 & 8.0 & ---  & 1.16\tablenotemark{b} & 10.93\tablenotemark{b} &  0.93\tablenotemark{b}                           & 10.88\tablenotemark{b} & 10.82\tablenotemark{b} & 10.85\tablenotemark{b} \\
Public Stack & all & 13 & 175\tablenotemark{c} & $\la214$ & --- & 8.7 & ---  & 1.37\tablenotemark{c} & 10.88\tablenotemark{c} &  ---  & --- & --- & 10.76\tablenotemark{c} \\
\tableline
\end{tabular}
\tablecomments{Column (1): GMASS ID from C08. Column (2): Redshift measured with pPXF. Column (3): 1$\sigma$ error in the velocity alignment. Column (4): Virial prediction for the velocity dispersion $\sigma^2_{\rm e}=G M_{\rm pop}/(5\re)$, corrected to a 1$\times$1 arcsec$^2$ aperture. Column (5): Measured galaxy velocity dispersion. Column (6): Error on $\sigma_\star$.  Column (7): $S/N$  per 60 \kms\ spectral pixel, computed from the pPXF fit residuals to the GMASS spectrum. Column (8): Circularized \re\ from the PSF-deconvolved MGE model. Column (9): \re\ in kpc. Column (10): Total $U$-band luminosity from the MGE model. Column (11): $U$-band mass-to-light ratio from the Jeans dynamical model. Column (12): Total mass from the Jeans model. Column (13): Virial estimate of the total mass. Column (14): Stellar population estimate of the mass from C08, using the models of \citet{Maraston2005} normalized for a \citet{Chabrier2003} IMF.}
\tablenotetext{a}{The spectrum of this galaxy was included in the New Stack. These values were computed by adopting for the galaxy the $\sigma_\star$ of the stacked spectrum of \refsec{sec:kinematics_stacked_16_20}.}
\tablenotetext{b}{Weighted mean $\langle u\rangle=\sum_j[{u_j (S/N)_j^2}]/\sum_j{(S/N)_j^2}$ of the quantities $u_j$ for the 7 galaxies included in the New Stack (\refsec{sec:kinematics_stacked_16_20}).}
\tablenotetext{c}{Weighted mean of the quantities for all 13 galaxies in the Public Stack (\refsec{sec:kinematics_stacked}).}
\label{tab1}
\end{table*}

\begin{figure}
\centering
\includegraphics[width=0.8\columnwidth]{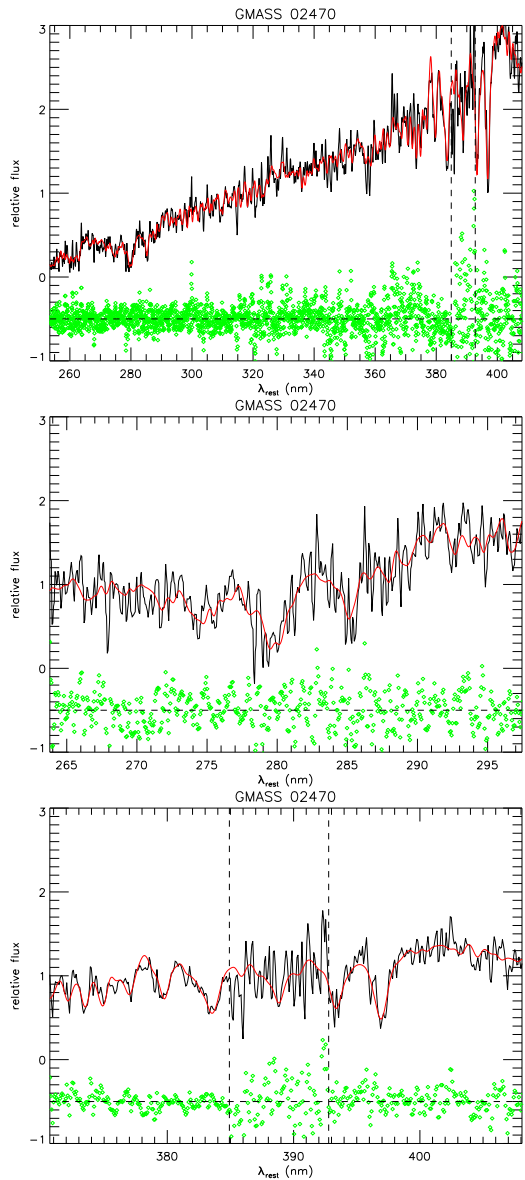}
\caption{Kinematics extraction for GMASS 2470 ($z\approx1.4$). The panels show the pPXF fits to the full spectral range (Top Panel), the blue (Middle Panel) and the red one (Bottom Panel) respectively. In each panel the black line is the observed spectrum, the red one is the best fitting template and the green diamonds are the residuals (arbitrarily shifted). The two vertical dashed lines indicate the spectral region excluded from the fit due to high noise due to sky lines.}
\label{fig:02470_ppxf}
\end{figure}

Some sharp absorptions are required for a reliable kinematics extraction. In the restframe wavelength range of interest (250--400 nm) the only significant ones are the \ion{Mg}{2} doublet (280 nm), \ion{Mg}{1} (285 nm) on the blue side, the \ion{Ca}{2} H and K ($\sim395$ nm) absorptions and a blend of \ion{Fe}{1} and \ion{Mg}{1} (384 nm) on the red side.

At $z\ga1.5$ the red spectral features fall outside our observed red range of 1000 nm and the kinematics relies on the \ion{Mg}{2} doublet and \ion{Mg}{1} absorptions. Sufficient $S/N$ is required for robust measurements at these redshifts. For this we could only measure reliable $\sigma$ for two $z\approx1.4$ individual galaxies, where the red features could be included in the fit. For all $\sigma$ determinations we used the Penalized Pixel-Fitting method\footnote{http://www-astro.physics.ox.ac.uk/$\sim$mxc/idl/\label{fn:ppxf}} (pPXF; \citealt{Cappellari2004}) with the 33 templates of \citet{Munari2005}, including additive polynomials, to correct for residual template mismatch or sky subtraction errors, and multiplicative polynomials, to correct possible spectral calibration errors. We verified the stability of our results with different degrees between 1--4 for the two sets of polynomials. In the fits 4--8 of the 33 templates were selected by pPXF to reproduce the spectrum.

The spectrum with the highest mean $S/N\approx8$ is GMASS 2470 (table~1 of C08). We measured $\sigma_{\rm obs}$ (before correcting for instrumental resolution) for three wavelength ranges (\reffig{fig:02470_ppxf}): (i) The full range (255--405 nm): $\sigma_{\rm obs}=192\pm13$; (ii) The blue range (264--297 nm): $\sigma_{\rm obs}=198\pm28$; (iii) The red range (371--408 nm): $\sigma_{\rm obs}=159\pm18$. The three results are consistent within the relative error bars, giving confidence in the adopted approach. We found in general no trend with wavelength so we adopt as standard value the one measured for the full spectral range, which as expected has smaller errors. The galaxy stellar dispersion $\sigma_\star$ is:
\begin{equation}
    \sigma_\star=\sqrt{\sigma_{\rm obs}^2-\sigma_{\rm instr}^2}.
\end{equation}
The values and errors for this galaxy and for GMASS 2239 are given in Table~\ref{tab1}.

\subsection{Public GMASS Stacked spectrum}
\label{sec:kinematics_stacked}

We applied the same approach of \refsec{sec:kinematics_individual} to measure $\sigma_\star$ from the public GMASS spectrum, obtained by coadding the individual normalized spectra of 13 ETG within $1.4\la z \la2.0$, for an unprecedented total integration time of 480-h and a mean $S/N\approx9$ (see fig.~4 of C08). We derived $\sigma_{\rm obs}=287\pm20$ \kms. Our value and error agree with the determination $\sigma_{\rm obs}=\sqrt{57^2+267^2}=273\pm20$ \kms\ (with high-resolution UV stars) performed on the {\em same} spectrum by \citet{Cenarro2009}. Those authors used empirical stellar template spectra, so this agreement validates our approach of using synthetic templates in the UV.

An additional step is needed to estimate the characteristic $\sigma_\star$ of the galaxies in the stack. In fact the spectra had to be shifted to the rest frame wavelength before coaddition. The measured redshift can be written as
\begin{equation}
1+z=(1+z_{\rm true})\times(1+V/c),
\end{equation}
where $z_{\rm true}$ is the true galaxy redshift, $V$ is the velocity shift due to an error in $z$ and $c$ is the speed of light. If the galaxies had all identical spectra and the redshift errors $\Delta z$ were normally distributed, the stacking would introduce an additional Gaussian velocity broadening
\begin{equation}
\sigma_{\rm stack}\equiv\Delta V\approx\Delta z\, c/(1+z)
\end{equation}
in the coadded spectrum. Assuming all broadening functions to be Gaussian, the dispersion of the individual galaxies could be recovered using
\begin{equation}
    \sigma_\star=\sqrt{\sigma_{\rm obs}^2-\sigma_{\rm instr}^2-\sigma_{\rm stack}^2}.
    \label{eq:stack}
\end{equation}
In the case of the public GMASS spectrum, which was {\em not} intended for kinematics measurements, the individual $z$ were measured via cross correlation and the smallest $\sigma_{\rm stack}\approx141$ \kms\ (M.~Mignoli private communication). One can then only derive an upper limit to the typical dispersion in the stack $\sigma_\star\la214$ \kms. This limit is smaller than the $\sigma_\star$ derived by \citet{Cenarro2009} as they incorrectly assumed $\sigma_{\rm stack}$ to be negligible.

\subsection{New stacked spectrum at $1.6\la z \la 2.0$}
\label{sec:kinematics_stacked_16_20}

\begin{figure}
\centering
\includegraphics[width=0.8\columnwidth]{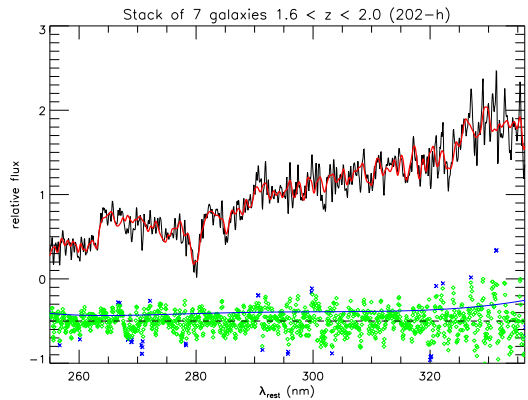}
\caption{Kinematics from stacked spectrum of \refsec{sec:kinematics_stacked_16_20}. In each panel the black line is the observed spectrum, the red one is the best fitting template and the green diamonds are the residuals (arbitrarily shifted). The blue crosses indicate pixels automatically excluded from the fit. The solid blue line indicates the estimated 1$\sigma$ noise.}
\label{fig:stacked_ppxf}
\end{figure}

We re-measured $z$ of all 13 GMASS galaxies with pPXF and give redshifts and errors in Table~\ref{tab1}. The new average velocity-error becomes $\sigma_{\rm stack}\approx30$ \kms, which is negligible with respect to the expected dispersions. We verified the reliability of our errors by measuring $z$ of the individual exposures of the same galaxy.

After excluding the two spectra of \refsec{sec:kinematics_individual}, to maximize the $S/N$ we constructed a stacked spectrum from the 7 remaining GMASS spectra with $S/N\ge2.5$. We normalized the spectra in the 260--310 nm wavelength range before coaddition, not to bias the kinematics towards the brightest galaxies. The measured $\sigma_\star=202\pm23$ \kms\ for the stack, corrected with \refeq{eq:stack}, agrees with the weighted average $\langle\sigma_{\rm pred}\rangle=205$ \kms\ of the virial predictions (Table~\ref{tab1}) for the galaxies in the stack. Although we do not trust the individual measured $\sigma_\star$ values for each low-$S/N$ spectrum in the stack, and do not give them in this paper, they are also not inconsistent with $\sigma_{\rm pred}$ and span the same range of values. The $\sigma_\star$ of the two individual galaxies of \refsec{sec:kinematics_individual} also agrees with $\sigma_{\rm pred}$. In Table~\ref{tab1} and in what follows we adopt the $\sigma_\star$ from the stack as representative of the $\sigma_\star$ of each of the 7 galaxies in the stack. This is {\em not} correct for each individual case, but only in an {\em average} sense.

\section{Dynamical models}

\subsection{Jeans modeling}

\begin{figure}
\centering
\includegraphics[width=\columnwidth]{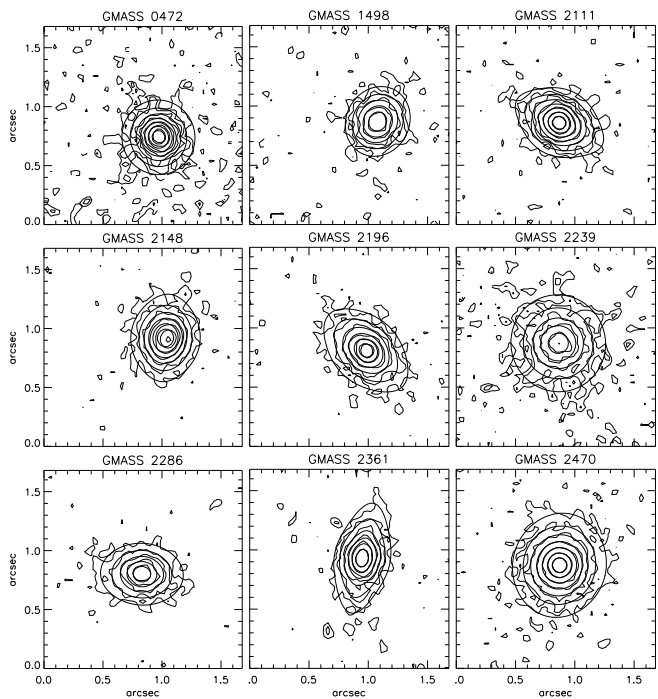}
\caption{MGE models for ten GMASS galaxies for which we constructed dynamical models. The contours of the observed HST/ACS/F850LP surface brightness are overlaid to an MGE model of their surface brightness, convolved with the ACS PSF. Contours are spaced in 0.5 mag arcsec$^{-2}$ intervals.}
\label{fig:mge}
\end{figure}

The $\sigma_\star$ we measured for the GMASS galaxies in \refsec{sec:dispersion} can be used to determine their dynamical masses. As the galaxies have half-light radii $\re\la0\farcs25$, while the spectra aperture and seeing have size of $\sim1\arcsec$ one may need significant corrections to the virial formalism used at low redshift to measure masses. One may estimate corrections using spherical Sersic dynamical models based on the Jeans equations \citep{diSeregoAlighieri2005}. However these models cannot describe well all ETGs, especially when they have disks and may rotate significantly. For this reason \citet{vanDerMarel2007} and \citet{vanderWel2008} used axisymmetric Jeans dynamical models of individual galaxies to take the surface brightness and possible rotation, as well as PSF and aperture, directly into account when measuring masses at high redshift. This is the approach we also use here.

We adopt a Multi-Gaussian Expansion (MGE) \citep{Emsellem1994} to parameterize the HST/ACS/F850LP ($z$-band) surface brightness of the GMASS galaxies (\reffig{fig:mge}), while taking the ACS PSF into account, using the software\footnotemark[\ref{fn:ppxf}] of \citet{Cappellari2002mge}. The following expression was used to $K$-correct the MGE parameters from observed countrate ($C_z$), in counts s$^{-1}$ per ACS pixel, into a restframe Johnson $U$-band surface brightness in mag arcsec$^{-2}$
\begin{equation}
    \mu_U=-2.5\log\left[\frac{C_z\times f_{850}\times (1+z)^5}{f_U(A0V)\times p^2}\right].
\end{equation}
Here $f_{850}=1.51\times10^{-19}$ ergs s$^{-1}$ cm$^{-2}$ \AA$^{-1}$ is the latest inverse sensitivity of the F850LP filter,\footnote{http://www.stsci.edu/hst/acs/analysis/zeropoints} $f_U(A0V)=4.28\times10^{-9}$ ergs s$^{-1}$ cm$^{-2}$ \AA$^{-1}$ is the zero point of the Johnson $U$-band, $p=0\farcs03$ is the dithered pixels size of the GOODS images.\footnote{http://archive.stsci.edu/pub/hlsp/goods/v2/h\_goods\_v2.0\_rdm.html}  We include both the $(1+z)^4$ bolometric dimming of the surface brightness and a factor $(1+z)$ due to the redshifting of the bandwidth. The formula is accurate at $z\approx1.4$, where the ACS/F850LP band is de-redshifted into the $U$-band. At larger redshifts we applied a small extra $K$-correction inferred from the stacked GMASS spectrum.

For each redshift we placed the models at the corresponding angular diameter distance $D_A$. We computed a prediction for the velocity second moment ($V^2_{\rm rms}=V^2+\sigma^2$) inside a 1\arcsec\ square aperture, with a 1\arcsec\ seeing FWHM, assuming semi-isotropy ($\beta_z=0$) and axisymmetry, for a constant $(M/L)_U=1$, using equation~(28) of the Jeans Anisotropic MGE (JAM)\footnotemark[\ref{fn:ppxf}] method of \citet{Cappellari2008}. We assumed an intermediate inclination $i=60^\circ$ for all galaxies, but the results do not change more than 5\% for an edge on inclination ($i=90^\circ$). The dynamical $M/L$ of each galaxy is then given by $(M/L)_{\rm Jeans}=(\sigma_\star/V_{\rm rms})^2$ (Table~\ref{tab1}). The $(M/L)_{\rm Jeans}$ decreases by 5\% by assuming in the models the largest radial anisotropy $\beta_z=0.5$ observed in nearby galaxies.

Not all 7 galaxies included in the stack are expected to have the same $\sigma_\star=202$ \kms\ we measured. Some can be higher and some lower than this average. If the virial predictions $\sigma_{\rm pred}$ were correct, the fact that generally $\sigma_{\rm pred}<202$ \kms\ suggests the quoted masses and $M/L$ are mostly overestimated.

\subsection{Virial and population masses}

\begin{figure}
\centering
\includegraphics[width=0.7\columnwidth]{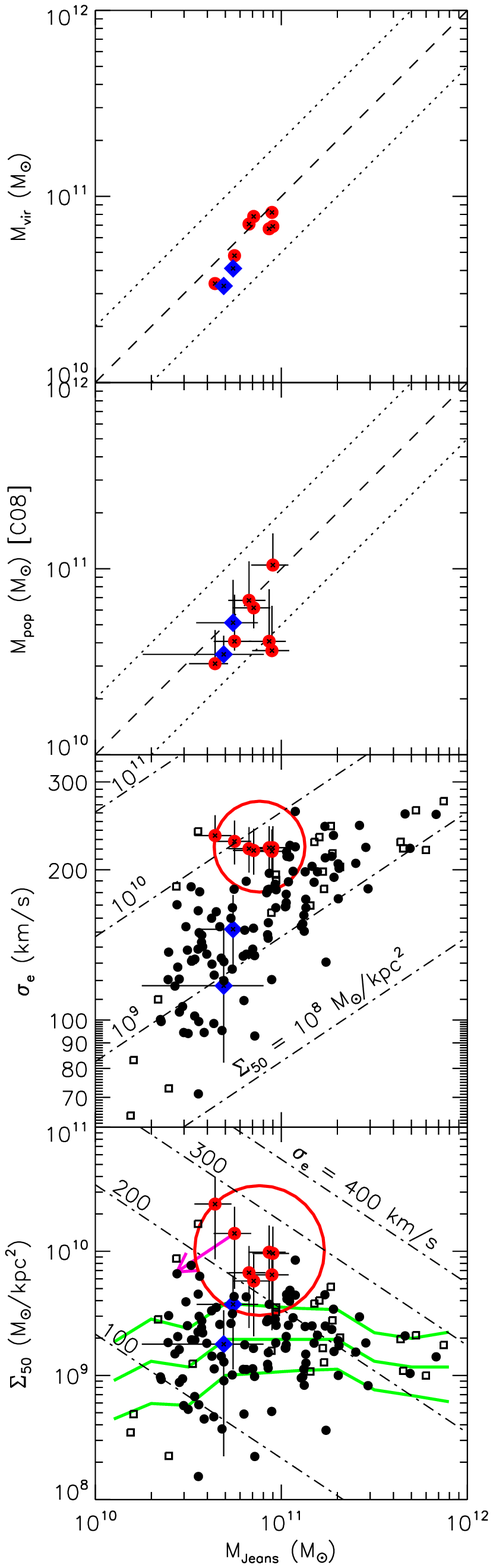}
\caption{\footnotesize {\em Top Panel:} Comparison between mass determinations via dynamical models $M_{\rm Jeans}$ and virial masses $M_{\rm vir}$. Blue diamonds are galaxies with individually measured $\sigma_\star$ (\refsec{sec:kinematics_individual}), while red circles are galaxies for which we assumed the $\sigma_\star$ of the stacked spectrum (\refsec{sec:kinematics_stacked_16_20}). The dashed line indicates equality, while the dotted lines correspond to a factor 2$\times$ difference.
{\em Second Panel:} As in the top panel, for a comparison with stellar population masses $M_{\rm pop}$. The error bars in the latter span the ranges of estimates using the three different population codes presented in C08. The symbols correspond to the \citet{Maraston2005} models.
{\em Third Panel:} Comparison between the GMASS \se, the values for the Coma sample (black filled circles) and dynamical models of local ETGs (black open squares). The large red open circle indicates the weighted average value for the seven galaxies in the New Stack (Table~\ref{tab1}). The dash-dotted lines are virial predictions of mass surface density $\Sigma_{50}$ inside \re.
{\em Bottom Panel:} As in the third panel, for $\Sigma_{50}$. We adopted errors of 30\% on \re. The dash-dotted lines are virial predictions of $\sigma_e$. The solid green lines are the values and errors from \citet{Shen2003}. The magenta arrow represents a 30\% decrease of $\sigma_\star$.}
\label{fig:mass_comparison}
\end{figure}

In the top panel of \reffig{fig:mass_comparison} we compare the dynamical mass obtained from the JAM models $M_{\rm Jeans}=L_U\times(M/L)_{\rm Jeans}$ to the virial mass $M_{\rm vir}=5.0\times\re\se^2/G$, where the scaling factor was calibrated using dynamical models and integral-field data of local ETGs, for \se\ measured within 1\re\ \citep{Cappellari2006}. We estimate \se\ by increasing $\sigma_\star$ from the measured $1\arcsec\times1\arcsec$ aperture to a 1\re\ circular aperture using equation~(1) of \citet{Cappellari2006}. Our \re\ values were determined in a non-parametric way from circularized MGE models ($\{\sigma_j,q'_j\}\leftarrow\{\sigma_j\sqrt{q'_j},1\}$) which preserve the luminosity and peak surface brightness of each Gaussian. With constant ellipticity this corresponds to the circularized radius $\re=\sqrt{a b}$ of the ellipse enclosing half of the analytically-derived MGE galaxy light. Our values agree (except for GMASS 2196) with the determination via Sersic profiles fits of C08 within their quoted 20\% errors (estimated via simulations).

There is a general agreement between the JAM and virial estimate, but in median the latter is $\sim30$\% lower. As both values are based on the same $\sigma_\star$, the difference must be attributed to an underestimation of \re\ and/or to non-homology in the profiles. This may be due to the low $S/N$ caused by cosmological surface brightness dimming \citep{Mancini2009}.
The JAM approach has the important advantage over the virial one that it robustly recovers the $M/L$ even when non-homology is important or the outer parts of the profiles are lost in the noise. Considering a test model with an $I(R)\propto\exp(-k R^{1/4})$ surface brightness profile truncated at 1\re, we still recovered the true $M/L$ to 1\% with JAM, but the $M/L$ was underestimated by 26\% with the virial approach.

In the second panel of \reffig{fig:mass_comparison} we compare $M_{\rm Jeans}$ to the mass determination $M_{\rm pop}$ based on stellar population models and 11 photometric bands of C08. The values are in agreement within the rather large uncertainty in both quantities. The agreement may improve when considering the possible underestimation of $\sigma_\star$ for some galaxies in the stack. This shows that mass errors are $\la2\times$ when detailed photometric information is available. It also confirms the result of C08 that ETGs at $z\sim2$ are consistent with a passive evolution in the past $t\sim1$ Gyr and indicates a formation redshift $z\sim3$. Any significant star formation activity would have dramatically lowered the dynamical $(M/L)_U$ which scales linearly with time in the age-sensitive $U$-band. This is in agreement and extends to $z\sim2$ previous dynamical studies of $M/L$ evolution based on the the Fundamental Plane at $z\sim1$ \citep{vanDokkum2003,Gebhardt2003,vandeVen2003,vanderWel2004,Treu2005,diSeregoAlighieri2005,diSeregoAlighieri2006,Jorgensen2006}.

The $M_{\rm pop}$ values are based on the \citet{Chabrier2003} Initial Mass Function (IMF). Adopting a Salpeter IMF would increase $M_{\rm pop}$ by 70\%, making in most cases $M_{\rm Jeans}<M_{\rm pop}$ for these high redshift galaxies. Similarly with a straight Salpeter IMF the $(M/L)_{\rm pop}$ ratio of local ETGs would be about twice the value derived from dynamical modeling \citep{Renzini2005,Cappellari2006}, which requires instead a bottom-light IMF such as in the case of \citet{Kroupa2001} or Chabrier's IMFs. Therefore, it appears that the dynamical modeling of both low redshift and high redshift ETGs requires a bottom-light IMF.

\section{Discussion}

We have measured the stellar velocity dispersion $\sigma$, from individual and stacked spectra, and have constructed detailed dynamical models, of 9 early type galaxies (ETGs) from the GMASS sample (C08) in the redshift range $1.4\la z\la2.0$. The agreement between the dynamical masses and the ones previously derived via population models by C08 indicates that an overestimation of the mass can not explain the high density discovered by previous works.

If high-$z$ ETG are indeed denser that local ones, they should have a higher $\sigma$ and surface mass density $\Sigma_{50}\equiv M_{\rm Jeans}/(2\pi\re^2)$ within \re\ at given dynamical mass \citep{Toft2007,vanDokkum2008}. To test this fact, in the bottom two panels of \reffig{fig:mass_comparison}, we compare the measurements for our GMASS galaxies to a sample of ETGs in the Coma cluster \citep{Jorgensen1999,Jorgensen2006} and to dynamical models of local ETGs \citep{Cappellari2006}, which use the same modeling technique as this paper. We also compare with the density derived on SDSS galaxies by \citet{Shen2003}, increased by 30\% to account for the fact that the population masses using a Kroupa IMF on average underestimate the dynamical mass of massive early-type galaxies (e.g.\ fig.~17 of \citealt{Cappellari2006}). We find that our two $z\approx1.4$ galaxies have $\sigma$ and $\Sigma_{50}$ consistent with the ones of local ETG (as shown in C08). However the galaxies in the stacked spectrum at $1.6<z<2.0$ have {\em on average} the $\sigma$ and $\Sigma_{50}$ of the most dense local ETGs.

This paper illustrates the limits of what can be achieved on the study of the dynamics of ETG with the current generation of telescopes. It emphasize the usefulness of stacking technique to infer the dynamics of selected classes of galaxies. Much progress along these lines could be obtained with massively multi-object spectrographs on the future generations of 30--40 m telescopes like the E-ELT. Access to an atmosphere-free near-infrared wavelength range, as soon available on James Webb Space Telescope, would dramatically improve the kinematics determination in ETGs at $z\ga2$ by bringing the rich set of optical absorption lines into the observable domain.

\section*{Acknowledgments}
We are grateful to Inger J{\o}rgensen for providing the virial parameters for the Coma galaxies. MC acknowledges support from a STFC Advanced Fellowship (PP/D005574/1). ED thanks ANR-08-JCJC-0008 funding.

%\bibliography{cappellari2009_gmass_dispersion}

\end{document}